\documentstyle[12pt]{article}
\begin{document}
\bibliographystyle{[baron.wp]aip}
\baselineskip = 22.2pt plus 0.2pt minus 0.2pt
\lineskip = 22.2pt plus 0.2pt minus 0.2pt

\renewcommand{\thesection}{\arabic{section}.}
\newcommand{\lton}{\stackrel{\large <}{\sim}}
\newcommand{\gton}{\stackrel{\large >}{\sim}}
\newcommand{\leqn}{\stackrel{\large <}{=}}
\newcommand{\geqn}{\stackrel{\large >}{=}}
\newcommand{\procent}{^o/_o}
\newcommand{\raw}{\rightarrow}
\newcommand{\beq}{\begin{equation}}
\newcommand{\bdm}{\begin{displaymath}}
\newcommand{\beqar}{\begin{eqnarray}}
\newcommand{\beqas}{\begin{eqnarray*}}
\newcommand{\eeq}[1]{\label{#1} \end{equation}}
\newcommand{\edm}{\end{displaymath}}
\newcommand{\eeqar}[1]{\label{#1} \end{eqnarray}}
\newcommand{\eeqas}{\end{eqnarray*}}
\newcommand{\gfm}{{\rm GeV/Fm}^3}
\newcommand{\epsi}{\vec{\epsilon}}
\thispagestyle{empty}
\pagestyle{myheadings}
\markboth
{\it  Ray, Shamanna and Kuo / Liquid-gas phase transition in Nuclear matter }
{\it  Ray, Shamanna and Kuo / Liquid-gas phase transition in Nuclear matter }

\begin{center}
{\bf Isospin Lattice-Gas Model and \\
Liquid-Gas Phase Transition in Asymmetric Nuclear Matter  }

\vskip 1cm
S. Ray \\
Saha Institute of Nuclear Physics \\
Calcutta 700 064 \\

\vskip .2cm

J. Shamanna, T.T.S. Kuo \\
Physics Department, State University of New York at Stony Brook,
Stony Brook, NY 11794-3800, USA
\\[2ex]

\today \\[3ex]
\end{center}
\hrule
\vskip .3cm
\begin{abstract}
 An isospin lattice-gas model, which is a spin-1 Ising model, is employed to
investigate the liquid-gas phase transition in asymmetric nuclear matter.
We consider nuclear matter as a lattice where each lattice site can be
either empty or occupied by a proton or a neutron, with a nearest-neighbor
interaction among the nucleons.
 With the Bragg-Williams mean field approximation, we calculate
 various thermodynamic properties of nuclear matter
for different densities and different proton-neutron
asymmetry parameter $s$.  Our model exhibits liquid-gas phase transition
below a critical temperature $T_c$, and predicts
a monotonic decreasing of $T_c$ as the magnitude of $s$ is 
increased.
The dependence of  the nuclear matter isotherms on the asymmetry parameter
$s$ is discussed.

\vskip .3cm
\hrule
\end{abstract}
\vfill
\eject

\section{Introduction}

The study of the equation of state (EOS) of asymmetric nuclear
matter has been, in the last few years, a subject of renewed interest
particularly in connection with astrophysical problems\cite{bck85,bgh85},
such as supernovae explosions and the evolution of neutron stars.
The EOS for the latter samples a range of densities
and isospin asymmetry which are different from those for supernovae.
Thus these two physical systems provide a unique
laboratory where the EOS of nuclear matter can
be critically investigated and give the possibility of obtaining
related but not identical information about the EOS.

Several laboratories have  studied the nuclear matter EOS by way of heavy-ion
collisions, such as the recent $^{197}Au$+$^{197}Au$ reaction \cite{gsi95}
investigated at GSI-Darmstadt. The proton-neutron ratio in  $^{197}Au$
is approximately 2/3. Clearly, the nuclear matter formed via such reactions
should be significantly asymmetric, lending further support to the need
for carrying out theoretical studies of the EOS for asymmetric nuclear matter.

In the past, asymmetric nuclear matter calculations have been done using
standard many-body methods, such as the HF calculations using the
Skyrme interactions\cite{jmz83,sylk86} and the Gogny
interaction\cite{szs90}. Brueckner-Hartree-Fock calculations for asymmetric
nuclear matter, using realistic
nucleon-nucleon (NN) interactions have also been performed
\cite{bl91,bkl94,swk92}.

We would like to consider a different approach for the EOS of assymmetric
nuclear matter. In statistical physics, phase transitions in
extended systems are generally studied using lattice (Ising) models.
For instance, the lattice gas model of
Lee and Yang \cite{le52}, where a gas with one type of atoms is mapped
into an Ising model for spin-$\frac{1}{2}$ particles, has
successfully described the liquid-gas phase transitions for atomic systems.
It should be of interest to explore the feasibilty of describing the
nuclear-matter EOS using lattice-gas model. Indeed, several authors
\cite{sk90,srks94,pdg195,pdg395,krss96} have employed lattice-gas
models to study the liquid-gas phase transition in symmetric
nuclear matter. In this paper we present a
lattice-gas model for investigating the liquid-gas phase transitions in
asymmetric nuclear matter.

\newpage

\section{Formulation}
We revisit the lattice hamiltonian for
nuclear matter proposed by Kuo et.al.\cite{sk90,krss96},
\beq
H_{int}= - \sum_{<ij>} J_{ij}\tau_{zi} \tau_{zj} -h \sum_{i} \tau_{zi}
\eeq{1}
The summation index $<ij>$ implies nearest neighbor interaction and $h$ is
some applied external field.
$J_{ij}$ are interaction strength parameters : $J_s$  for symmetric pairs
(pp and nn) and $J_d$ for asymmetric type pairs (pn pairs).
The $J_{ij}$ have the property
\beqar
J_{ij} = \cases{J_{ij} &~~~~ for nearest neighbor distance~=~$a$ \cr
	         \infty &~~~~ for neighbor distance~=~0 \cr
		 0 &~~~~ otherwise }
\eeqar{2}
\noindent where $a$ is the lattice spacing.
The isospin $\tau_z$ is
\beqar
\tau_z = \cases{-1 &~~~~ for neutron \cr
	          0 &~~~~ for vacancy \cr
		  1 &~~~~ for proton }
\eeqar{3}
If $N_{++},N_{--}$ and $N_{+-}$ represent respectively the
nearest neighbor pairs of proton-proton, neutron-neutron and
proton-neutron, the interaction hamiltonian may be written as,
\beq
H_{int}= -J_s(N_{++}+N_{--})+J_d N_{+-} -h(N_{+}-N_{-})
\eeq{4}
There is no interaction between vacancy and nucleon.

\section{Thermodynamics}
Let $N$ denote the total number of lattice sites,
and $N_+,N_- ,N_0$ the number of proton, neutron
and vacancy sites respectively. The relative emptiness ($r$)
and proton-neutron asymmetry parameter ($s$) are defined as,
\beq
r=\frac{N_0}{N},\;\;s=\frac{N_+ - N_-}{N},\;\; N=N_+ +N_- + N_0
\eeq{5}
The nuclear-matter density $\rho$
is proportional to $(1-r)$ in our model.

For atomic systems, where the spin-$\frac{1}{2}$ lattice gas models have
been used with remarkable success in describing phase transitions,
Huang \cite{hu63} suggests adding the ideal gas pressure to the lattice
gas grand potential in order to describe the isotherms. The ideal gas
pressure comes from the kinetic energy.
This motivated us to include a kinetic-energy term in our hamiltonian itself,
namely
\beq
H_{gas}= -J_s(N_{++}+N_{--})+J_d N_{+-}+ N\kappa (1-r)^{5/3} - Nhs
\eeq{6}
where we have assumed the kinetic energy per particle to be proportional
to $\rho ^{2/3}$ as per the Fermi gas model.
$\kappa$ is a constant which will be discussed later.
From the above hamiltonian we can write down the partition function. For the
 evaluation of the partition function, we have employed
the Bragg-Williams mean field approximation\cite{hu63}, namely
\beq
\frac{N_{+}^2}{N^2}\simeq\frac{N_{++}}{N \gamma  /2},\;\;
\frac{N_{-}^2}{N^2}\simeq
\frac{N_{--}}{ N \gamma /2},\;\; \frac{N_{0}^2}{N^2}\simeq\frac{N_{00}}
 { N \gamma /2}
\eeq{7}
where  $\gamma$ denotes the number of nearest neighbors to any given site,
and $ N\gamma /2$ is the total number of pairs. For three dimensional
simple cubic lattice, $\gamma =6$.
If one counts the number of
bonds between a site and all its nearest neighbors, then for (say) all
the proton sites one obtains
\beq
\gamma N_+ = 2N_{++} +N_{+-} +N_{+0}
\eeq{8}
\noindent and similarly for the neutron and vacant sites.

Using the above we can eliminate $N_{+-}$ as,
\beq
N_{+-}=\gamma (N_+ +N_-) -(N_{++}+N_{--}) +N_{00} -\gamma N/2
\eeq{9}
The hamiltonian of eq. (6) can then be rewritten as
\beq
H_{gas}(r,s,N) = -C_{1}Ns^{2}-C_{2}N(1-r)^2+N\kappa(1-r)^{5/3}-hNs
\eeq{10}
where
\beq
C_{1} =\frac {(J_s+J_d) \gamma }{4},\;C_{2}=\frac {(J_s-J_d)\gamma }{4}
\eeq{11}
Unlike the symmetric case, $s$ is now a parameter and not a variable.
The canonical partition function is therefore given by
\beq
Q_{gas}= \sum_{r=0}^{1}~g(r,s,N)~e^{(-\beta H_{gas})}
\eeq{12}
\noindent where the multiplicity factor \cite{will} $g(r,s,N)$ is given by
\beq
g(r,s,N)=\frac{N!}{N_{0}!N_{+}!N_{-}!} = \frac{N!}
{(Nr)![\frac{N}{2}(1-r+s)]![\frac{N}{2}(1-r-s)]!}
\eeq{13}
To study the system at varying densities, we
need to work with the grand partition function with a chemical potential
$\lambda$
\beqar
Q_G&=& \sum_{r}~ g(r,s,N)~z^{(N_{+}+N_{-})}~e^{(-\beta H_{gas})} \nonumber \\
&=& \sum_{r}~ g(r,s,N)~z^{(N_{+}+N_{-})}~e^{(-\beta {H}_{gas})}
\eeqar{14}
\noindent where $z=e^{\beta \lambda}$ is the fugacity.

\vskip .45cm
In the thermodynamic limit ($N\rightarrow\infty$), the sum in eq. (14)
can be replaced by its most dominant term
\cite{will} (assuming the dominant term to be
non-degenerate). We have verified that this approximation is
acceptable by numerically checking the relative size of the
different leading order terms. Using Stirling's formula one obtains
\beqar
\ln Q_{G} &=& \beta C_{1}Ns^2 + \beta C_{2} N (1-r)^2 +\beta h N s
+ \beta \lambda N (1-r) - \beta \kappa N(1-r)^{5/3} \nonumber \\
& & - N[\frac{(1-r+s)}{2}\ln(1-r+s)
+ \frac{(1-r-s)}{2}\ln(1-r-s) \nonumber \\
& & + r \ln r -(1-r)\ln 2]
\eeqar{15}

\noindent with the  extremum condition
\beqar
\frac{\partial (\ln Q_{G})}{\partial r} & \equiv & -\frac{N}{2}
\ln [ \frac{4r^2}{(1-r)^{2}-s^{2}}]- 2 \beta NC_{2}(1-r)
+ \; \frac{5}{3} \beta \kappa N(1-r)^{2/3}  \nonumber \\
& & + \beta N\lambda = 0
\eeqar{16}
In a previous work \cite{krss96} it was shown that $h = 0$ and $s=0$ results
in a global maximum for this term. In the present case however
we allow $s$ to vary parametrically. Factoring out the common terms
eq. (16) is written as
\beq
k_B T \ln [ \frac{4r^{2}}{{1-r}^{2}-s^{2}}] + 2C_{2}(1-r)- \frac{5}{3}
\kappa (1-r)^{2/3}+ \lambda = 0.
\eeq{17}
A graphical analysis of the above equation reveals the existence of a
condensed phase.

\section{Phase transition and critical temperature}
Let us consider the $\lambda$=0 case first. In this case we
rewrite eq. (17)
as
\beq
\chi (r,T) = k_BT\cdot f(r,T)-g(r)=0
\eeq{18}
\noindent with
\beqas
f(r) &\equiv & \frac{1}{2} \ln  [\frac{4r^2}{(1-r)^2-s^2}]  \\
g(r) &\equiv & \frac{5 \kappa}{3} (1-r)^{2/3} -2 C_{2}(1-r)
\eeqas
In the region $r \in [0,1]$, $k_BT \cdot f(r)$ is a monotonically increasing,
unbounded function of $r$.
It has one point of inflection where it intersects the $r$ axis and this
is at $r =\frac{1}{3}$ when $s=0$.
In the same domain $g(r)$ however, is a bounded function with a negative
curvature and one maximum in the region $r\in[0,1]$. Thus there is a
possibility of $g(r)$ intersecting with $k_BT \cdot f(r)$ more than once below a
suitable temperature $T_c$.
$k_BT \cdot f(r)$ intersects the $r$ axis at $r=\frac{-1 +
\sqrt{1+3(1-s^2)}}{3}$
which depends on $s$ but is independent of the value of $T$. On the
other hand $g(r)$ intersects
the $r$ axis at two points: One is at  $r=1$ which is independent of
the values of the parameters $C_2,\; \kappa,$ and $s$, and the second
point of intersection depends on $s$ and the ratio
$\alpha \equiv 5 \kappa/6 C_2$.
For a suitable choice of the parameter $\alpha$, the functions
 $k_BT \cdot f(r)$ and $g(r)$ intersect
the $r$ axis at a common point given by $r=\frac{-1 + \sqrt{1+3(1-s^2)}}{3}$.
The $T_{c}$ is then determined by that value of temperature for a given $s$
at which the curves $k_BT\cdot f(r)$ and $g(r)$ are tangential at their common
zero point (point of intersection with the $r$ axis).

In Fig. 1 we display some typical behaviours of $k_BT \cdot f(r)$ and $g(r)$.
The points of intersection of the two curves are shown for a family of
$k_BT \cdot f(r)$ and $g(r)$ with $k_BT=5,10,15$ MeV. This is done
for several values of the asymmetry parameter.

For $T < T_c$, the curves $k_BT \cdot f(r)$ and $g(r)$ have three
intersection points, as denoted by $A$, $B$, and $C$ for the $k_BT=5 $ MeV
case.
It is readily checked that the middle intersection point, i.e. $B$,
corresponds to a minimum of $\ln Q_G$ and hence it is not the physically
interesting solution that we are looking for.
The intersection points $A$ and $C$ are the physical solutions.

At some critical temperature $T_c$ the curves $k_BT_c \cdot f(r)$
and $g(r)$ are
tangent to each other at their common zero point. Above this critical temperature
$T_{c}$, the two curves intersect only once at their common point of
intersection. Below $T_{c}$ however, the slope of $k_BT \cdot f(r)$ at its zero point is
less than that of $g(r)$ and two more points of intersection appear which
represent the densities of the two different phases.
Thus the critical vacancy $r_0$ and the critical temperature $T_c$ are
determined by
\beqar
k_BT_c \cdot f(r_0) &=& g(r_0) \nonumber \\
k_BT_c \cdot f^{\prime}(r_0) &=& g^{\prime}(r_0)
\eeqar{19}
As the magnitude of the asymmetry parameter $s$ increases, the curve $T \cdot f(r)$
becomes steeper and multiple intersection with $g(r)$ becomes possible only
at lower temperatures. For pure neutron matter ($s=-1.0$) the critical
temperature from our model is equal to $0$ and there is only one phase as is
expected. (Fig. 3)

\section{Equation of state and isotherms}
The equation of state is obtained as
\beqar
P(\bar r,T) & \equiv & \frac{k_BT}{N} \ln Q_{G} \nonumber \\
&=& C_1s^2+hs+C_2(1-\bar r)^2+\lambda(1-\bar r)-\kappa(1-\bar r)^{5/3} \nonumber \\
& & -k_BT[\frac{(1-\bar r +s)}{2} \ln(1-\bar r+s)
+ \frac{(1-\bar r-s)}{2} \ln (1-\bar r-s) \nonumber \\
& & + \bar r \ln \bar r-(1-\bar r)\ln2]  \\
\frac{1}{v} & \equiv & \frac{z}{N}~\frac{\partial }{\partial z}
 \ln Q_G =(1-\bar{r})
\eeqar{21}
Note that the first two terms contribute just an additive constant
to the pressure for any given $s$.
We calculate the p-v isotherms using
eqs.(18), (20) and (21). Our results are shown in Figs. 4 to 6.

For a given temperature $T$ the isotherm is
obtained as a parametric plot of specific volume versus pressure with
$\lambda$ its generating parameter.
For $ T \geq T_c $ one gets a single smooth curve by varying
$\lambda$. The phase boundary corresponds to $\lambda = 0$. The isotherms to
the left of this boundary are obtained with positive $\lambda$, those to the
right with negative $\lambda$.
The figure suggests the existence of three regions below $T_{c}$,
the dense (small $r$, large $\rho$) liquid-like phase, the rare (large $r$, sma
ll $\rho$) gas-like phase and the coexistence phase in between.

For $T$ below $T_c$, it is of  interest to note that no
isotherms are obtained in the intermediate region, i.e. the coexistence
region.
The above is because when $T<T_c$, eq. (18) has no
solutions in the coexistence region.  
With the introduction of an
infinitesimal  $\lambda$ the system chooses one of the two values of
$\bar r$ (or $v=\frac{1}{(1-\bar r)}$) admissible for the given $T < T_c$,
depending on the sign of $\lambda$. This phenomena is reminiscent of the
spontaneous symmetry breaking in ferromagnetism.

The isotherms for different asymmetry parametered systems are
shown in Fig. 4, 5 and 6. These graphs are obtained by a very densely chosen
set of points where eq. (18) is solved and the corresponding pressure and
volume determined. To avoid falling into the trap
of seeing a smooth isotherm when it is not, we intentionally
chose not to use spline, or any higher polynomial fit.
A very densely packed set of points with short straight line connects
generated the isotherms. The absence of jaggedness that
can be seen is coming from the physics and calculation itself, and
not due to an artifact of any particular curve fitting technique.

\section{Analysis and results}
To have an attractive nearest-neighbor interaction,
$J_s >0$ and $J_d <0$. In this case  $C_2$ is positive.
The magnitude of $J_s$ and $J_d$ is comparable to the average
potential energy in nuclear matter, about $-40$ MeV. Hence
$C_2 \simeq 125$ MeV, recalling that $C_2$ has been defined in eq. (11).
The parameter $\kappa$ may be estimated from the average kinetic energy
given by the Fermi gas model.

Near the critical point, there is a subtle dependence of the
solutions of eq. (18)  on the ratio  $\alpha \equiv \frac{5\kappa}{6C_2}$.
For a choice of $\alpha =(2/3)^{1/3}$, the three intersection points
below $T_c$ ($A$, $B$ and $C$ of Fig. 1), all merge together at the
critical point as in the $s=0$ case\cite{krss96}.
Then the phase diagram near $T_c$ has the familiar smooth shape. For a
different choice of $\alpha$,  the merging would generally take place in two steps, first
involving two intersection points and then the third. This will lead to a phase
diagram with a {\it cusp} shape near the critical point, which seems
rather unconventional.
It would be interesting to investigate these unconventional phase boundaries
further for different values of $\alpha$.
For the present study however, we have chosen $\alpha$
so as to get a smooth phase
boundary. This choice of $\alpha$ is one for which the curves $k_BT \cdot f(r)$ and
$g(r)$ intersect the $r$ axis at a common point.
With this $\alpha$ and $C_2$=125, we have $\kappa$=131.037, which has been
used in obtaining the results presented in our Figs. 1 to 6.

For fixed $T$, the solutions of eq. (17) determine the $r$ values where
the grand partition function is maximum. Very similar
to what was observed in the $s=0$ case, this maximum
term is an overwhelmingly dominant one.
(Numerical simulation for finite lattices has shown that the magnitude of
the maximum term is typically a factor of $\sim 10^{80}$ greater than the other
terms.)
Hence the $r$ values given by eq. (18) are just the average values
(stricly speaking the modal value) of $r$,
denoted by $\bar r$, for the system at temperature $T$, asymmetry $s$
and $\lambda$=0 for which $\ln Q_{G}$ is maximum. The $\bar r$ values for
$\lambda \neq 0$ are given by the solution of eq. (17).

We find support of our earlier observation \cite{krss96} in our present
calculation as well, that a term of Fermi gas type kinetic energy
is needed for appropriate phase structure to appear.
If there is no kinetic energy term, $\kappa=0$ and hence $\alpha = 0$.
Then the requirement that at $T< T_c$, $g(r)$ intersect $T \cdot f(r)$ more
than once cannot be met, and our model would have no phase transitions.
\section*{Conclusion}
Our simple model exhibits the existence of a liquid gas phase transition
in asymmetric nuclear matter. The p-v isotherms
obtained by our model look surprisingly similar to those given by the
van der Waals theory, except for one crucial difference that for $T < T_c$
our isotherms do not have the metastable states in the coexistence region.
Hence with our model one does not need to determine the phase boundary
by way of a Maxwell construction.

The liquid-gas critical temperature
obtained is dependent on the asymmetry parameter $s$, $T_c$ decreases
with increasing neutron density as shown in Fig. 3. The value is around
$T_c = 17-20$ for typical values of $s$ which is fairly close to the
results given by earlier 
calculations\cite{jmz83,su87,bbfgl88,krss96}. 
The existence of a liquid-gas phase transition together with the
determination of its critical temperature on the basis of a
simple model that assumes only a phenomenological
two-body, nearest-neighbor interaction is quite remarkable.

\vfill\eject

\vfill\eject
\begin{center}
{\bf{FIGURE\,\,\,CAPTIONS}}
\end{center}
\begin{description}
\item[Fig.1] Graphical solution of eq.(18) for s=-0.2.
\item[Fig.2] Graphical solution of eq.(18) for s=-0.4.
\item[Fig.3] Critical Temperature versus asymmetry parameter.
\item[Fig.4] Nuclear matter p-v isotherms for s=-0.2.
\item[Fig.5] Nuclear matter p-v isotherms for s=-0.4.
\item[Fig.6] Nuclear matter p-v isotherms for s=-0.6.
\end{description}

\end{document}